\documentclass[aps,twocolumn,pre]{revtex4}

\usepackage{graphicx}
\usepackage{latexsym}
\usepackage{amsmath}
\usepackage{amsfonts}
\usepackage{amssymb}

\newcommand{\be}{\begin{equation}}
\newcommand{\ee}{\end{equation}}
\newcommand{\bea}{\begin{eqnarray}}
\newcommand{\eea}{\end{eqnarray}}
\newcommand{\gammac}{\gamma_{\mathrm{contact}}}

\begin{document}
\title{Annealed lattice animal model and Flory theory for the melt of non-concatenated rings: Towards the physics of crumpling}
\author{Alexander Y. Grosberg}
\email{ayg1@nyu.edu} \affiliation{Department of Physics and Center for Soft Matter Research, New York University, 4 Washington Place, New York, NY 10003 USA}


\date{\today}

\begin{abstract}
A Flory theory is constructed for a long polymer ring in a melt of unknotted and non-concatenated rings.  The theory assumes that the ring forms an effective annealed branched object and computes its primitive path.  It is shown that the primitive path follows self-avoiding statistics and is characterized by the corresponding Flory exponent of a polymer with excluded volume.  Based on that, it is shown that rings in the melt are compact objects with overall size proportional to their length raised to the $1/3$ power.  Furthermore, the contact probability exponent $\gammac$ is estimated, albeit by a poorly controlled approximation, with the result close to $1.1$ consistent with both numerical and experimental data.
\end{abstract}
\maketitle

\section{Introduction}

In recent years, significant attention has been paid to the problem of a dense system of long, unknotted, and non-concatenated rings, or melt of rings for brevity \cite{Cates_Deutsch_1986, Obukhov_Rubinstein_Duke_PRL_1994, Vilgis_Rings_1995, Muller96, Muller2000, Rubinstein_Nature_2008, Vettorel_Grosberg_Kremer_2009, Suzuki09, Vettorel_Developing_Entanglements, Melt_of_Rings_Statics_2011, Melt_of_Rings_Dynamics_2011, Linear_contaminants, Sakaue_Rings_Brief_PhysRevLett.106.167802, Sakaue_Rings_Detailed_PhysRevE.85.021806, Rubinstein_talk_in_Leiden, Comparing_Lattice_and_off-Lattice}. One source of excitement is the beautiful intellectual challenge, because rings question the most cherished idea of polymer topology, that of tubes \cite{DoiEdwards}.  Furthermore, unlike many other problems in polymer physics, this one does not have any known field theoretic equivalent and thus we lack any systematic method to calculate anything.  In more practical terms, rings seem to be the most promising start in making some special materials, with ultra-low elastic modulus \cite{Rubinstein_Nature_2008}. Last but not least, the problem appears to have direct relevance for chromatin folding, the question of genome architecture in the living cell (see, e.g., reviews \cite{Dekker_Chapter_7, Mirny_Chromosome_Res_Review, Mirny_CurrentOpinion_2012} and multitude of references therein).

In this context, chromosome conformation capture (3C and HiC) \cite{3C, HiC_Science_2009, HiC_for_mouse, HiC_drosophila_Sexton} experiments, as well as observations of chromosome territories \cite{Chromosome_territories_Cremer}, strongly rule out any equilibrium polymer conformation, and instead seem to agree with the crumpled (or fractal) globule model \cite{Crumpled_Globule_Model_1993}, in a variety of simulated versions \cite{HiC_Science_2009, Rosa_Everaers_PLOS_2008, Vettorel_Grosberg_Kremer_2009, Comparing_Lattice_and_off-Lattice}. Despite promising parallels between experiments and simulations, theoretical understanding of either real chromatin or any of the simulated models remains elusive. The challenge is to find theoretical insight into the interplay between high density and non-concatenation in a polymer system with quenched topology.  A melt of rings is arguably the simplest model which may help us to meet this challenge.

There are several questions one may want to answer regarding a melt of rings (assume each ring unknotted, and all rings mutually non-concatenated; assume also for brevity that the monomer size and Kuhn segment are equal to each other and taken to be the unit of length):

\noindent \textit{\textbf{Question 1:}} What is the overall size of the ring, $R$, and what is the power $\nu$ of its scaling with the ring length, $N$: $R \sim N^{\nu}$? The same index also describes the size of a subchain of length $s \ll N$, as $r(s) \sim s^{\nu}$.

\noindent \textit{\textbf{Question 2:}} What is the probability $P(s)$ that two monomers a distance $s \ll N$ apart along the chain meet in space? This is governed by another index $\gammac$ which seems independent of $\nu$: $P(s) \sim s^{-\gammac}$.

\noindent \textit{\textbf{Question 3:}} What is the ``primitive path'', or Cayley tree diameter, or spanning distance, or generation number $L$ (see \cite{Khokhlov_Nechaev_1985, Rubinstein_PRL_1986, Arya_PRE} and below) of one ring in the melt, and how does it scale with $N$: $L \sim N^{\rho}$? The same index $\rho$ also governs the behavior of the primitive path for a subchain: $p(s) \sim s^{\rho}$.

Historically, researchers have mostly concentrated on question (1) and index $\nu$.  Cates and Deutsch \cite{Cates_Deutsch_1986} developed a Flory theory which indicated $\nu = 2/5$. Brereton and Vilgis \cite{Vilgis_Rings_1995} employed the linking number as the only relevant topological invariant and obtained $\nu = (3 \pi - 1)/6\pi \approx 0.45$.  At the same time the idea of the crumpled globule \cite{crumpled1} implied that long polymers in a dense system cannot penetrate if reptation is suppressed, which means every ring in the melt of rings should be a collapsed object with $\nu = 1/3$.   Early computer simulations \cite{Muller96} seemed to agree with the $\nu=0.4$ prediction, but the latest data for much longer chains \cite{Vettorel_Grosberg_Kremer_2009, Suzuki09, Melt_of_Rings_Statics_2011} convincingly demonstrate a wide cross-over into a regime where the asymptotic index is below $0.4$ and consistent with $\nu=1/3$.  The $\nu=1/3$ scenario for the longest chains seems to be gaining popularity and is postulated as obvious in the recent works \cite{Sakaue_Rings_Brief_PhysRevLett.106.167802, Sakaue_Rings_Detailed_PhysRevE.85.021806, Rubinstein_talk_in_Leiden}.

Even less is known theoretically about other exponents.  With regard to $\gammac$, theoretical arguments to date have not gone beyond the mean field statement $\gammac = 3 \nu$, which, if $\nu = 1/3$, is impossible, because of a pathological divergence of the number of neighbors for one monomer.  Numerically \cite{Melt_of_Rings_Statics_2011, Comparing_Lattice_and_off-Lattice} and experimentally for chromatin \cite{HiC_Science_2009,HiC_for_mouse} $\gammac$ appears to be just slightly above unity, close to $1.1$ (between $1.05$ and $1.20$). And $\rho$ for rings was never a subject of any calculation, only some \textit{ad hoc} assumptions \cite{Obukhov_Rubinstein_Duke_PRL_1994, Rubinstein_talk_in_Leiden} or simulations \cite{Arya_PRE}.

In this work, we start from question (3) and construct a Flory theory to compute index $\rho$. This can be done using the ideas of the works \cite{Annealed_Branched,Disordered_UFN}. The Flory theory we construct suggests that the primitive path is organized in space as a self-avoiding walk. It becomes possible to argue then that this proves $\nu = 1/3$. Finally, these considerations along with the ideas of the works \cite{Duplantier_arbitrary_network_1989, Khokhlov_reactions} allow us to give a rough estimate for $\gammac$, which is found to be consistent with the results of both simulations and experiments.

\section{Ring size and spanning distance: a theory of indices $\nu$ and $\rho$}

The key idea of our approach is the assumption that every ring in the unconcatinated melt is, roughly, double folded to form, on sufficiently large scales, a branched structure like a lattice animal.  The most important aspect here is the fact that branched structure is annealed, subject to thermal motion and equilibration. This is definitely true for an ideal ring placed in the lattice of topological obstacles or in a gel \cite{Khokhlov_Nechaev_1985, Rubinstein_PRL_1986, Nechaev_Semenov_Koleva_1987, Obukhov_Rubinstein_Duke_PRL_1994}.  In this case, annealed branched structure is naturally mapped on the Cayley tree, and its size there, $L$, is called a primitive path or spanning distance.  Physically, $L$ describes the degree of branching: if $L$ is as large as $N/2$, our ring is a double folded linear chain, whereas if $L$ is as small as $\sim \ln N$ it is like a dendrimer.  We imagine that some sort of self-consistent Cayley tree representation is also valid for every ring in the melt, as each ring is squeezed in a self-consistent manner by the surrounding rings \cite{Obukhov_Rubinstein_Duke_PRL_1994, Rubinstein_talk_in_Leiden}. This assumption is far from trivial.  Although implicitly suggested by the amoeba-like dynamics of rings in the melt \cite{Melt_of_Rings_Dynamics_2011}, branched structures are not visible to the naked eye in simulated conformations \cite{Melt_of_Rings_Statics_2011}, and they do not show up in the version of the primitive path analysis implemented in the work \cite{Melt_of_Rings_Statics_2011}.  These two arguments, however important, do not prove the absence of an underlying tree-like structure hidden from the eye. The algorithm of detection for such a hidden tree is not known, and it is the focus of future work. Short of solid proof one way or another, a pragmatic strategy is to assume that some annealed branched structure is there, examine the consequences, and then return to the main assumption at the end. That is what we do in this work.

Note that $L \sim N^{\rho}$ and $R \sim N^{\nu}$ imply $R \sim L^{\nu/\rho}$, which means $\nu/\rho$ is an index which governs the shape of the primitive path ``backbone'' in real space. This allows us to formulate weak but simple bounds on the possible values of $\rho$. Since $R \leq L \leq N$, we have
\be \nu \leq \rho \leq 1 \ . \label{eq:weak_constraints} \ee

To gain a grasp on the index $\rho$ and compute it, consider the following Flory theory. There are two competing terms in the free energy for the branched object representing rings in the melt. First, there is the usual term $R^2/L b^2$ which resists stretching, except it is stretching of the primitive path rather than the polymer itself, thus there is $L$ instead of $N$ in the denominator, and ``monomer size'' $b$ will be defined later. Second, there is a similar term which penalizes for stretching of the polymer along the Cayley tree or, in other words, penalizes for insufficient branching of the lattice animal: $L^2/N$. Both terms are of similar structure, except $L$ plays the role of ``polymer length'' in the former and the role of ``spatial size'' in the latter. To make it a bit more accurate, we have to take care of the factors involving the entanglement length $N_e$, because what maps on the Cayley tree is the chain of blobs, $N_e$ monomers and $N_e^{1/2}$ size each. Therefore, we just have to replace $L \to L/N_e^{1/2}$, $b \to N_e^{1/2}$, and $N \to N/N_e$. This produces free energy (in units of thermal energy $k_BT$)
\be \frac{\Delta F}{k_B T} \sim \frac{R^2}{L N_e^{1/2} } + \frac{L^2 }{N} \ . \label{eq:Flory_free_energy} \ee
Apart from $N_e$, such a free energy expression was derived for the annealed branched polymer in the work \cite{Annealed_Branched}. The variational parameter here is $L$.  To understand the nature of formula (\ref{eq:Flory_free_energy}), it is useful to forget for a moment about $N_e$ and remember that we will eventually show compactness, $R \sim N^{1/3}$; if we replace $N \to R^3$ in the last term, we obtain \textit{exactly} the regular Flory free energy for a linear chain -- except usually the variational parameter is $R$, while in our case it is $L$. It is thus not surprising that our Flory theory produces self-avoiding statistics.

Optimization of $\Delta F$ yields the equilibrium value of $L$
\be L \sim \frac{R^{2/3} N^{1/3}}{N_e^{1/6}} = N_e^{1/2} \left( \frac{R}{N^{1/3}N_e^{1/6}} \right)^{2/3}\left( \frac{N}{N_e} \right)^{5/9} \ \ee
and minimal free energy
\be \frac{\Delta F}{k_BT} \sim \frac{R^{4/3}}{N^{1/3} N_e^{1/3}} = \left(\frac{R}{N^{1/3} N_e^{1/6}} \right)^{4/3} \left( \frac{N}{N_e} \right)^{1/9} \ . \label{eq:Equilibrium_free_energy_depends_on_R} \ee

This free energy (\ref{eq:Equilibrium_free_energy_depends_on_R}) depends on the overall spatial size $R$ monotonously and favors small $R$. For a branched polymer in a good solvent \cite{Annealed_Branched}, there is an additional usual repulsive second virial term in free energy ($\sim N^2/R^3$) which favors swelling above Gaussian size.  In the melt situation, there is no such term in the free energy and no such physical factor.  In the ideally incompressible system (e.g., fully occupied lattice) there are no other contributions to the free energy apart from purely entropic ones in formulae (\ref{eq:Flory_free_energy}) or (\ref{eq:Equilibrium_free_energy_depends_on_R}). In such an ideal system there would be a strict inequality $R \geq R_{\mathrm{dense}}$, and the minimum of the free energy (\ref{eq:Equilibrium_free_energy_depends_on_R}) would exactly saturate this bound.  For a real system, we can do marginally better by incorporating an extra term in the free energy which properly diverges as $R$ approaches its minimal value, as was done in \cite{Sakaue_Rings_Brief_PhysRevLett.106.167802, Sakaue_Rings_Detailed_PhysRevE.85.021806}. To scaling accuracy, this of course adds nothing new to the statement that the free energy (\ref{eq:Equilibrium_free_energy_depends_on_R}) drives the system to the minimal possible value of $R$ consistent with the self-excluded volume constraint for every ring.  This is actually a well known situation for branched systems in $d<4$ as well as for linear polymers in $d<2$ \cite{Pincus_Stockmayer_1983}.

Here, we should realize that $R_{\mathrm{dense}}$ is somewhat larger than naive $N^{1/3}$, because of $N_e$: in each blob, the self-density is $\sim N_e^{-1/2}$, therefore, $N_e^{1/2}$ rings must overlap, which means $R_{\mathrm{dense}} \sim N^{1/3} N_e^{1/6}$.  The role of this overlap between rings due to $N_e$ and its relation to the so-called Kavassalis-Noolandi criteria \cite{Kavassalis_Noolandi_PhysRevLett.59.2674} were recently emphasized by Rubinstein \cite{Rubinstein_talk_in_Leiden}. Thus, we obtain equilibrium values of both overall ring size $R$ and its spanning diameter $L$, or corresponding indices $\nu$ and $\rho$:
\begin{subequations}
\begin{align}
 \label{eq:equilibrium_size}
 R & \sim \left\{ \begin{array}{lcr} N^{1/2} & \text{for} & N  \ll  N_e \\  N^{1/3} N_e^{1/6} & \text{for} & N \gg N_e \end{array} \right. \ \ \text{or} \ \ \nu = 1/3  \ , \\
\label{eq:equilibrium_spanning_Flory}
   L & \sim \left\{ \begin{array}{lcr} N^{1/2} & \text{for} & N  \ll  N_e \\ N^{5/9} N_e^{-1/18} & \text{for} & N  \gg  N_e \end{array} \right. \ \ \text{or} \ \ \rho = 5/9  \ .
\end{align}
\end{subequations}
Together, these two imply
\be \label{eq:size_of_spanning}
 R \sim \left\{ \begin{array}{lcr} L & \text{for} & L  \ll  N_e^{1/2} \\  L^{3/5} N_e^{1/5} & \text{for} & L  \gg  N_e^{1/2} \end{array} \right. \ \ \text{or} \ \ \frac{\nu}{\rho} = 3/5  \  . \ee
This result is very suggestive.  Based on it, one can argue that the statistics which governs placement of the primitive path (like a backbone) in space is that of a self-avoiding walk.  For this, of course, $3/5$ is just an approximate value of the critical exponent which in general is the well known Flory index $\nu_F \approx 0.588 \approx 3/5$; therefore, the result (\ref{eq:equilibrium_size},\ref{eq:equilibrium_spanning_Flory}) is
only a mean field approximation to the relations which are hereby hypothesized to be exact:
\be \nu = 1/3 \ , \ \ \rho = 1/3\nu_F \ ,  \ \ \nu/\rho = \nu_F \ .
\ee
Of course, it satisfies constraints (\ref{eq:weak_constraints}).

The $\nu = 1/3$ result (lower line of Eq. (\ref{eq:equilibrium_size})) agrees nicely with simulations for the longest chains \cite{Vettorel_Grosberg_Kremer_2009, Suzuki09, Melt_of_Rings_Statics_2011}. Trivially, the $\nu=1/2$ for short rings (upper line of Eq. (\ref{eq:equilibrium_size})) also agrees with simulations. The present theory has nothing to say about the cross-over region. Simulation results indicate that the cross-over is rather broad, roughly between $0.3 N_e$ and $20 N_e$. This motivates the idea that there might be an intermediate asymptotics, possibly corresponding to $\nu = 0.4$. For instance, two characteristic length scales (instead of one $N_e$) were postulated in theoretical works \cite{Sakaue_Rings_Brief_PhysRevLett.106.167802, Sakaue_Rings_Detailed_PhysRevE.85.021806}, with an intermediate $N^{4/9}$ regime (this result was not derived in \cite{Sakaue_Rings_Brief_PhysRevLett.106.167802, Sakaue_Rings_Detailed_PhysRevE.85.021806}, but it follows from their theory).  Unfortunately, the existence of a second independent length scale is not borne out in any simulation or experimental data. For instance, as shown in \cite{Linear_contaminants, Comparing_Lattice_and_off-Lattice}, data from many works by different groups using different models collapse on a single master curve when $R$ is plotted against $N/N_e$, with the entanglement length $N_e$ independently measured for every model, from rheology and/or primitive path analysis for linear chains. This collapse is a powerful indication that there is no other length scale in the problem independent of $N_e$.  The uniqueness of $N_e$ as the only length scale is also consistent with the Kavassalis-Noolandi criteria \cite{Kavassalis_Noolandi_PhysRevLett.59.2674}. The present theory, in agreement with all existing simulations and experiments, assumes no second length scale and, accordingly, does not produce any intermediate asymptotics. Instead, one can easily engineer many interpolation formulae smoothly connecting the two asymptotics indicated in equation (\ref{eq:equilibrium_size}), and some of them fit well with the universal master curve from the works \cite{Linear_contaminants, Comparing_Lattice_and_off-Lattice}.  We do not show here any of these interpolation cross-over formulae because they have no physical justification, and we do not want to obscure the fact that understanding of the cross-over region is lacking. This is in contrast to long rings, for which the above theory seems to give a rather complete picture.

Another explanation of the above theory was suggested by R.Everaers \cite{Everaers_Private_Communication}.  It addresses a possible dissatisfaction with the fact that $\nu=1/3$ was obtained above as a result of inequality imposed by the excluded volume, and interactions were not explicitly taken into account.  Let us add a virial interaction term of some general order $p+1$ to the variational free energy (\ref{eq:Flory_free_energy}):
\be \frac{\Delta F}{k_B T} \sim \frac{R^2}{L N_e^{1/2} } + \frac{L^2 }{N} + \frac{1}{N} N_e^{pd/2}\left(\frac{N/N_e}{R^d} \right)^{p+1} R^d \ . \label{eq:Everaers_free_energy} \ee
Here, the $1/N$ pre-factor accounts for the screening effect in the melt \cite{Isaacson_Lubensky_1980}, which is done for rings in the same way as for linear chains, because screening is entirely due to the translational entropy.  Interacting units are blobs of $\sim N_e$ monomers, $N/N_e$ is their number, and $N_e^{pd/2}$ is their virial coefficient.  Optimizing free energy (\ref{eq:Everaers_free_energy}) with respect to both $L$ and $R$ (in any order) yields $R \sim N^{\nu(p,d)}$ with $\nu(p,d) = (1+3p)/(4 + 3 d p)$.  In $d \leq 4$, we have $\nu(p,d) \leq 1/d$, which means virial term of any finite order $p+1$ is not strong enough to prevent collapse to the un-physically small size smaller than $N^{1/d}$.  However, at $p \to \infty$ we do obtain $R \sim N^{1/d}$; in particular, in $d=3$ we recover the complete result (\ref{eq:equilibrium_size}).

\section{Contact probability: a speculative estimate of the index $\gammac$}

Having now achieved a good understanding of exponents $\nu$ and $\rho$ for long rings, we have to discuss question (2) and index $\gammac$.  Unfortunately, we found so far only a rather speculative estimate of this index.  Nevertheless, presenting it seems useful, as at least it illustrates the nature of the problem.

The contact probability for rings is usually considered for monomer pairs with contour distance $s \ll N/2$. It is in this range that the power law $P(s) \sim s^{-\gammac}$ is valid, otherwise correlations through the $N-s$ arm of the ring are also important. More generally, for any $s$, assuming there is no other scale for $s$ except $N$, we should be able to write $P(s) = s^{-\gammac} \phi(s/N)$, where the scaling function $\phi(x)$ is such that $\phi(x) \simeq 1$ when $x \ll 1$, and $\phi(x) \simeq \left( \frac{x}{1-x} \right)^{\gammac}$ when $x \to 1$. More importantly for us now is the statement that $\phi(1/2)$ is just a number of order unity. Therefore, the contact probability between two opposite points on the ring scales as
\be P_{\mathrm{opposite}} = P(N/2) \sim N^{-\gammac} \ . \label{eq:contact_of_opposite} \ee

\begin{figure}[h]
\centering
\includegraphics[width=0.35\textwidth]{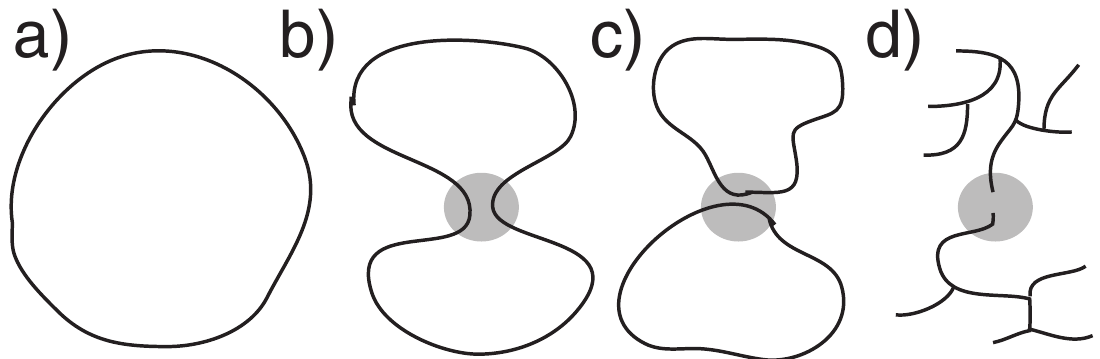}
\caption{Explanation for equation (\ref{eq:Khokhlov_formula}). }
\label{fig:Loop_with_Contact}
\end{figure}

This result is convenient for finding $\gammac$ because the contact probability for two opposite points of the ring can be determined from the relation (similar to the one suggested for linear chains by Khokhlov \cite{Khokhlov_reactions})
\be P_{\mathrm{opposite}} \sim \frac{1}{R^3} \frac{\Omega_{\rho} (N)}{\Omega_{\rho}^2(N/2)} \ , \label{eq:Khokhlov_formula} \ee
where $\Omega_{\rho}(N)$ is the number of conformations for a lattice animal of $N$ monomers whose branching degree is characterized by index $\rho$. This relation can be understood in the following way.  The probability $P_{\mathrm{opposite}}$ is proportional to the number of conformations for a $N$-long ring (fig. \ref{fig:Loop_with_Contact}a) with two opposite points glued together (fig. \ref{fig:Loop_with_Contact}b), or, equivalently, for two rings, $N/2$ long each, glued together in one point (fig. \ref{fig:Loop_with_Contact}c).  Since each of these $N/2$-rings is effectively a generalized lattice animal with known index $\rho$, what we need is the probability of ``fusion'' between ends of two $N/2$-animals provided that they overlap and, therefore, share the volume $\sim R^3$ (fig. \ref{fig:Loop_with_Contact}d). Since $\Omega_{\rho} (N)/\Omega_{\rho}^2(N/2)$ is the volume where one half-animal is positioned with respect to the other such that their ends are within one bond length, the ratio of this volume over $R^3$ is the requisite probability -- which is the formula (\ref{eq:Khokhlov_formula}).

Number of conformations $\Omega$ is usually expressed as
\be \Omega_{\rho} = z^{N} N^{\gamma_{\rho} -1 } \ ,
\label{eq:number_of_states} \ee
where $z$ is the non-universal fugacity, and $\gamma_{\rho}$ is a universal critical exponent.  We should emphasize that different types of lattice animals belong to different universality classes \cite{Annealed_Branched}, so $\gamma_{\rho}$ does depend on $\rho$. Plugging this general formula for $\Omega$ into equation (\ref{eq:Khokhlov_formula}), keeping only the $N$-dependent factors, and also using $R \sim N^{\nu}$, we arrive at
\be P_{\mathrm{opposite}} \sim \frac{1}{N^{3\nu}} \frac{z^{N} N^{\gamma_{\rho} -1}}{z^{2(N/2)} (N/2)^{2 (\gamma_{\rho} - 1)}} \sim N^{-\gamma_{\rho}+1-3\nu} \ . \ee
Since this must be $\sim N^{-\gammac}$, and since $3 \nu = 1$ we get
\be \gammac = \gamma_{\rho} -1 + 3 \nu = \gamma_{\rho} \ . \ee
This is already an important result, for it relates the contact exponent $\gammac$ to the much more studied exponent of ``usual'' $\gamma$, albeit for a peculiar object (i.e., an annealed branched polymer).  This result explain also the difficulty of finding the index $\gammac$: index $\nu$ in various polymer problems can usually be satisfactorily estimated using Flory theory, while no analog of Flory theory exists for index $\gamma$.  At the same time, $\epsilon$-expansion approach is also not applicable to the problem of rings, because topological constrains for rings do not exist in $d>3$ and, accordingly, no field theoretic formulation exists for the problem.

What is known in terms of $\gamma_{\rho}$ is only $\gamma_1$, which is the case of linear chains. In this case, $\gamma_1 \approx 1.16156 \pm 0.0020$ according to $\varepsilon$-expansion \cite{Accurate_gamma_RG}, $\gamma_1 \approx 1.1575 \pm 0.0006$ according to high precision Monte Carlo \cite{Accurate_gamma_PhysRevE.57.R1215}. Here we use very limited accuracy, so $\gamma_1 \approx 1.16 \approx 7/6$.

We can try to consider the situation perturbatively, assuming that $\rho$ is very close to unity, or $1-\rho \ll 1$.  If $\rho$ is very close to unity, then the polymer overall is almost linear, with backbone of length $N^{\rho}$ close to $N$ and with short side chains.  This object as a whole is close to a linear polymer and the non-exponential factor in its number of distinct conformations can be estimated as $L^{\gamma_1 - 1} = \left(N^{\rho} \right)^{\gamma_1-1} = N^{\rho (\gamma_1 - 1)}$.  Thus
\be \gamma_{\rho} \approx 1+ \rho (\gamma_1 - 1) \ . \ee
This estimate is surely imperfect, its accuracy is not controlled; one can call it even speculative.  This is the weakest part of our estimate of $\gammac$.  Nevertheless, taking this estimate for granted and making the next bold step of assuming that $5/9$ is close enough to unity, we arrive at
\be \gammac = \gamma_{5/9} \approx 1+ \frac{5}{9} \left( \gamma_1 - 1 \right) \approx \frac{59}{54} \approx 1.09 \ . \label{eq:theoretical_gamma} \ee

This result is consistent with the numerical and experimental data. Experimentally, $\gammac$ is reported to be $1.08$ for human cells \cite{HiC_Science_2009} and $1.05$ for mouse \cite{HiC_for_mouse}, although it is unclear what are the associated error bars. Numerically, direct measurements of contact probability yield $\gammac \approx 1.2$ for both lattice and off-lattice models \cite{Comparing_Lattice_and_off-Lattice}, while indirect measurements based on surface roughness give for $\gammac$ values from $1.03$ based on surface contacts counting to $1.07$ based on static structure factor fitting \cite{Melt_of_Rings_Statics_2011}. At this time we can only conclude that theoretical result (\ref{eq:theoretical_gamma}) is in the right ballpark. Further development of both numerical and theoretical understanding will be necessary to clarify the matter.

\section{Discussion and conclusion}

Our results on indices $\nu$ and $\rho$ shed light on the following single molecule question.  Consider a single unknotted loop densely squeezed in a volume of the size $N^{1/3}$. What are the subchain sizes $r(s)$ in this case?  Is there any difference between confinement in a cavity with smooth walls and confinement in a 3d torus, i.e., in periodic boundary conditions with the same volume $N$?  Either of these two situations is different from the melt of rings, because the confinement size is smaller than in the melt: $N^{1/3} < N^{1/3} N_e^{1/6}$.  Trivially, we expect Gaussian behavior $r(s) \sim s^{1/2}$ at $s<N_e$ crossing over to $r(s) \sim (s/N_e)^{\nu} N_e^{1/2}$ at $s>N_e$, but this ``crumpled'' regime for $r(s)$ becomes problematic when subchain size reaches overall confinement size.  This happens at the scale $s^{\ast}$ such that $\left( s^{\ast}/N_e \right)^{\nu} N_e^{1/2} = N^{\nu}$.  We expect that $r(s)$ will continue to grow in the case of periodic boundary conditions, meaning that the subchain will wind more than once around the torus.  By contrast, subchain size $r(s)$ must saturate at $s>s^{\ast}$ for the cavity case.  Thus: 
\bea r_{\mathrm{torus}}(s) & \simeq & \left\{ \begin{array}{ccc}
                           s^{1/2} & \mathrm{for} & s<N_e \\
                           s^{1/3} N_e^{1/6} & \mathrm{for} & s>N_e 
                        \end{array}
   \right. \\
   r_{\mathrm{cavity}}(s) & \simeq & \left\{ \begin{array}{ccc}
                           s^{1/2} & \mathrm{for} & s<N_e \\
                           s^{1/3} N_e^{1/6} & \mathrm{for} & N N_e^{-1/2} >s>N_e \\
                           N^{1/3} & \mathrm{for} & N> s > N N_e^{-1/2}
                        \end{array}
   \right.  \ . \label{eq:subchain_size_three_regimes} \eea
In the cavity case, the window for the ``crumpled'' $s^{\nu}$ regime exists only as long
as $s^{\ast} > N_e$, which means
\be N \gg N_e^{1/2\nu} = N_e^{3/2} \ . \label{eq:condition} \ee
Numerical experience suggests that a strong inequality is required here.  The condition
(\ref{eq:condition}), when re-written as $N^{2/3} \gg N_e$, allows for an interesting insight:
$N^{2/3}$ in the equilibrium globule is the length of the chain which reaches as a Gaussian
random walk from one end of the globule to the other.  The condition requires that such a segment
be long enough to experience topological constraints of its own.  In this regard it would be
important to understand the relation between entanglement length $N_e$ and random knotting length
$N_0$.  So far we operate here under the tacit assumption that these two are of the same order,
but it is generally not known (see, however, \cite{Milner_relation_Ne_N0}).

Another noteworthy corollary of our results has to do with 2d melt of ``unconcatenated'' rings.  In 2d, the analog of unconcatenation is the condition that no ring is located inside any other ring.  In this case, the annealed branched object representation of the ring is an obvious rigorous fact.  We can easily reproduce our calculations for this case to show that rings in this case are, of course, compact, with $\nu = 1/2$, while their spanning distance obeys $R \sim L^{3/4}$, i.e., in 2d it is a self-avoiding walk, as in 3d.

Rubinstein \cite{Rubinstein_talk_in_Leiden}, based on the \textit{ad hoc} assumption $\nu/\rho=1/2$ (which means that the primitive path behaves as a Gaussian rather than self-avoiding random walk) and $\nu=1/3$, explored many dynamical properties using the methods similar to his paper \cite{Obukhov_Rubinstein_Duke_PRL_1994}. It is straightforward to reproduce his arguments for the newly determined value of $\rho$, and the results agree very well with simulations \cite{Melt_of_Rings_Dynamics_2011}. These results will appear in a forthcoming publication \cite{Generalized_Lattice_Animals_Dynamics}.

At the end it is prudent to return to the beginning and discuss the central assumption of this note, that of the annealed branched structure for the rings in the melt.  As mentioned in the beginning, we adopted here a pragmatic strategy to assume the existence of annealed branched structure and to examine the consequences. This approach itself can be traced back to the paper \cite{Arya_PRE}, where the authors simulated a branched system under formally (`by hand') imposed strong confinement in both real ($R$) and Cayley tree ($L$) spaces. We in the present work extend the approach much farther in that our Flory theory is self-consistent, it defines (to the scaling accuracy) both $R$ and $L$ from the physical conditions of dense packing and topological non-concatenation.  The result is nothing short of encouraging because it yields for the first time: a proof of compact statistics for every ring ($\nu = 1/3$), a proof of the relation between the contact probability exponent $\gammac$ and the traditional exponent $\gamma$, a reasonable (albeit based on an uncontrolled approximation) theoretical estimate of $\gammac$, a non-trivial index $\rho$ having to do with self-avoiding walk statistics, and a good description of rings dynamics \cite{Rubinstein_talk_in_Leiden, Generalized_Lattice_Animals_Dynamics}.

To conclude, it was shown that every ring in the non-concatenated melt is compact and is described by a primitive path which follows self-avoiding statistics.  This insight made possible the calculation of the contact exponent which was found to be in agreement with numerical and real experiments.

\textit{Added note:}  When this paper was already written down, an article \cite{RosaEveraers_Archive} appeared which delivers a very powerful numerical evidence in support of annealed lattice animal model of the rings in the melt.

Inspiration for this work came from a talk by Michael Rubinstein \cite{Rubinstein_talk_in_Leiden}.  The author acknowledges stimulating discussions with Ralf Everaers and Kurt Kremer, advice from Robijn Bruinsma and Yitzhak Rabin, and help in manuscript preparation from Jonathan Halverson.

\bibliography{Scaling_References}

\end{document}